\documentclass[10pt, conference, a4paper, twocolumns]{IEEEtran}

\usepackage{cite}
\usepackage[cmex10]{amsmath}
\usepackage{url}
\usepackage{graphicx}
\usepackage{color}
\usepackage{subcaption}
\usepackage[T1]{fontenc} % Pour avoir les lettres accentuees
\usepackage{stfloats}
\usepackage{longtable}
\usepackage{tabularx}
\usepackage{array}
\usepackage{cleveref}
\usepackage{multirow}
\usepackage{color}
\usepackage{enumitem}
\usepackage[table]{xcolor}

\definecolor{applegreen}{rgb}{0.55, 0.71, 0.0}

\title{How long delays impact TCP performance for a connectivity from Reunion Island ?}
\author{\IEEEauthorblockN{R\'ehan Noordally\IEEEauthorrefmark{1},
Xavier Nicolay\IEEEauthorrefmark{1},
Yassine Gangat\IEEEauthorrefmark{1}\IEEEauthorrefmark{2}%,
and Pascal Anelli\IEEEauthorrefmark{1}}\\
\IEEEauthorblockA{\IEEEauthorrefmark{1}Laboratoire d'Informatique et de Math\'ematiques}
\IEEEauthorblockA{\IEEEauthorrefmark{2}Laboratoire d'Energ\'etique, d'Electronique et Proc\'ed\'es}
\IEEEauthorblockA{University of Reunion Island, 15 Rue Ren\'e Cassin, 97490 Sainte Clotilde, France\\email : firstname.lastname@univ-reunion.fr}}

\begin{document}
\maketitle

\begin{abstract}
% state the problem
TCP is the protocol of transport the most used in the Internet and have a heavy-dependence on delay.
Reunion Island have a specific Internet connection, based on main links to France, located $10.000\ km$ away. As a result, the minimal delay between Reunion Island and France is around $180\ ms$.
In this paper, we will study TCP traces collected in Reunion Island University. The goal is to determine the metrics to study the impacts of long delays on TCP performance. 

% Reunion Island is a particular\yassine{"Faux ami à plusieurs sens. utilise le mot exact specific ? unusual ?"} case in terms of delay and routes. In this paper, \yassine{we want to hightligh} the impact of the particular\yassine{idem} connectivity on the characteristics and on the performance of the Internet traffic. 
%\rouge{NextComp $\Rightarrow$ 6 pages (including figures and references)}
\end{abstract}


\section{Introduction}
\label{sec:intro}
%% Evolution de l'Internet depuis ses débuts avec également évolutions des services. Une majorité de ces services utilisent TCP comme protocole de transport. TCP dépendant des délays. Réunion cas particulier vis à vis des délais du a du routage passant à 97\% des données analysés paqur la France métropolitaine. Est-ce que ce comportement particulier impact les performances de TCP ?
%\yassine{J'ai corrigé directement dans le texte qd c'est facilement compréhensible...}
%% Describe the problem
% Evolution of the Internet
% More and More TCP
% TCP is hardly relied to the delay
% Reunion Island have bad relationship with delay
% Performance of TCP in Reunion Island case.
Internet has evolved to become a large network where services will differ according to the user's need. Internet can be used, as it first purpose was, to send mail or use the \emph{World Wide Web (WWW)}. Now, on-line gaming, audio or videao discussion are currently used on the Internet. But all these new services require low delay for interactivity. %a high \emph{Quality of Service (QoS)} %with low delay. 
%To be sure that the QoS stay stable along the exchange, these services are mostly using the \emph{Transmission Control Protocol (TCP)} protocol. 
Due to the difficulty of using news transport protocols, these news services continue to use \emph{Transmission Control Protocol (TCP)}. This transport protocol is very sensitive to the variation of the delay. There are some geographic areas where the delay is higher than usual. In~\cite{Noordally2016}, the authors have shown that Reunion Island has a specific %particular\yassine{idem} 
behavior in terms of routing and delay. In this paper, we would investigate the impact of these characteristics on the performance of TCP. 
% Among the performance, we supervised the traffic and the different flows meet\yassine{"comprend pas"}\rouge{En plus des performances, on va effectuer une supervision du traffic et etudier les flux rencontrer selon differents criteres.}.

% State your contributions
After a selection of specific metrics, we have collected a total of 72 Gigabytes of data by passive metrology during one month using a listener. The analysis of the files revealed two evidences:
\begin{enumerate}
\item Up of 78\% of the traffic is going outside of the Island.
\item There is a strong presence of TCP in collected traffic. One most interesting finding is that HTTPS take near half of the service used by TCP.
\end{enumerate}

The remainder of this paper is organized as follows. The section~\ref{sec:description} 
explains the specific context of physical connectivity of Reunion Island.
%resumes the findings of the previous study related to the case . 
A description of the suitable metrics to analyze TCP traffic have been determined in the section~\ref{sec:metrics}, whereas the section~\ref{sec:tools} presents the tools used to capture and to analyze TCP traces. The section~\ref{sec:methodo} presents the methodology to collect traffic. Collected traffic is analyzed in section~\ref{sec:results}. We complete the paper with the overview of related work in section~\ref{sec:related} and our conclusions in section~\ref{sec:conclusion}.

%--------------------------------------
\section{Context}
\label{sec:description}
%% Presentation rapide de la situation de la réunion en terme de delay. Résultats de pascal en 2012, les miens en 2016 sur Krajsa et une nouvelle PDF réalisé sur 2016-2017.

% As seen on Figure~\ref{map:cables}, there exist two cables that connect Reunion Island to the Internet. The SAFE cable spans from Asia (India and Malaysia) to South Africa. The SAFE is extended by the SAT3-WASC cable which connects countries in West Africa to Europe. The LION and LION 2 cables connect the lower part of the Indian Ocean to a range of submarine cables in East Africa. This means that Reunion Island is connected to four landing points namely West Africa and Europe, East Africa and Asia. There is also an IXP~\cite{AXIS} in Reunion Island as well as in the nearby Mauritius Island. As a result, it should be enough to experience a fairly good connection toward theses areas and as well as the entire Internet.

\begin{figure}[ht!]
\centering
\includegraphics[width=0.98\columnwidth]{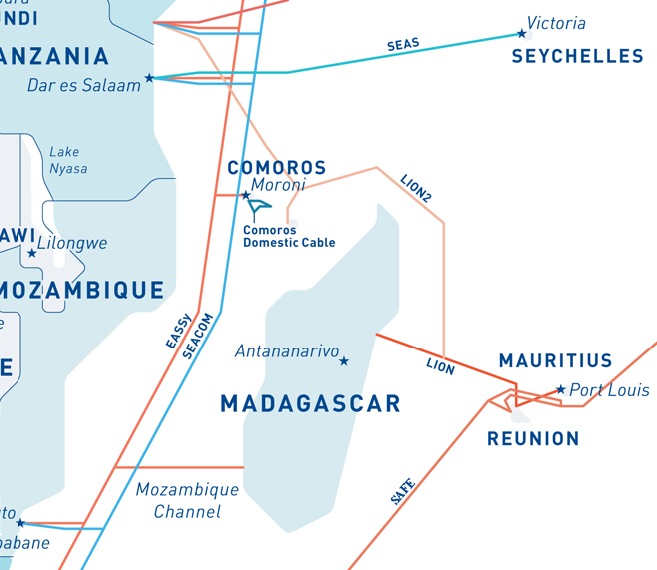}
\caption{Reunion Island Internet Connectivity}
\label{map:cables}
\end{figure}

% However, most of the residents of Reunion Island experience a slow Internet connection with frequent outages\footnote{According to the ISP, most of these outages are not induced by the local infrastructure.}.
%% \yassine{"Comme je t'avais dit la denrière fois, avant de commencer, tu devrais pitet parler de la situation géographique de la réunion et de sa situation "politique" ou "administrative" (comme étant parti de la france...)"}\rouge{pour moi c'est du hors-sujet. C'est pour ca que j'ai du mal a l'ecrire}\\
Reunion Island is a French overseas department located far away from France: in the Indian Ocean, between Madagascar and Mauritius. 
On the map~\ref{map:cables}, we can see that Reunion Island is connected to the Internet by two submarine cables. 
The first one is the \emph{South Africa Far East (SAFE)} spans from Asia (India and Malaysia) to South Africa. The SAFE is extended by the SAT3-WASC (\emph{South Africa Transit 3/West Africa Submarine Cable}) cable which connected West Africa to Europe. The~\emph{Lower Indian Ocean Network (LION)} and LION2 cables connect, as its name indicates, the lower part of the Indian Ocean to a range of submarine cables in East Africa. This means that Reunion Island is connected to four landing points namely West Africa and Europe, East Africa and Asia.
The author of~\cite{Anelli2012} compared the Internet latency observed from Reunion Island and Paris. 
\begin{figure*}[ht!]
	\begin{subfigure}[h]{\columnwidth}
		\includegraphics[width=0.98\columnwidth]{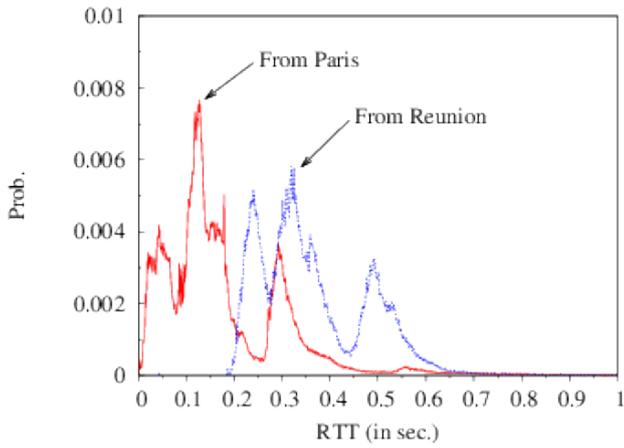}
		\caption{2012}
		\label{curve:2012}
  \end{subfigure}
  \hfill
  \begin{subfigure}[h]{\columnwidth}
  	\includegraphics[width=0.98\columnwidth]{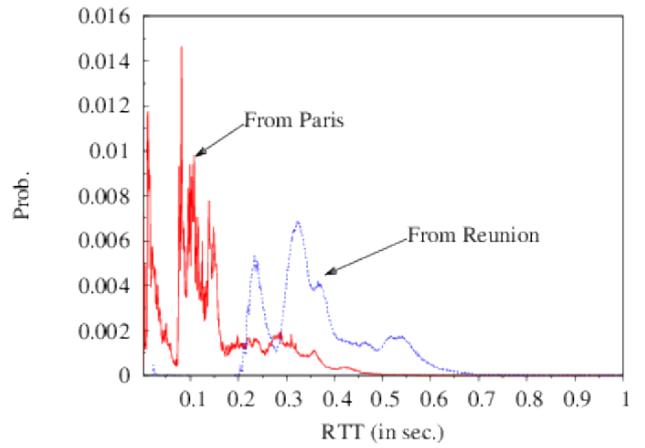}
		\caption{2016}
		\label{curve:2016}
  \end{subfigure}
  \caption{Comparison between Paris and Reunion Island access}
  \label{curve:ping}
\end{figure*}

The figures~\ref{curve:2012} and~\ref{curve:2016} represents the \emph{Probability Density Function (PDF)} of the \emph{Round-Trip Time (RTT)}. The RTT was evaluated with the \texttt{ping} command on a sample of IPv4 addresses distributed over the world. The first part colored in blue in~\ref{curve:ping} represents the density probability for delay from Reunion Island, while the red depicts the PDF from Paris. On the first figure, the two curves have the same trend with a $200\ ms$ shift.
To get a better comparison, we have updated this result with one new experiment. The figure~\ref{curve:2016} shown the updated results. The red curve shows a shorter delay than before, while the minimal delay from Reunion Island has increased. The delay from Reunion Island was smooth through time between the two experiments. 

This is one reason why we will study the impact of this specific % particular\yassine{idem} 
connectivity on the transport protocol performance.

%-----------------------------------------------------------------
\section{Metrics}
\label{sec:metrics}
%% Description des metriques choisies : Pourquoi ces métriques ? Quelles ont-elles de particulier pour qu'on les étudies.
In this section, we would describe each metric we have chosen to analyze TCP traffic. We have split the metrics in two categories: supervision and performance.

\subsection{Traffic Supervision}
\label{subsec:supervision}
%% Results supervision only here : Volume captured.
In this class of metrics, we would only have a view of the traffic exchange, as an ISP could have. %\yassine{"Si j'ai bien compris, il y a 3 métriques ds cette catégorie. Right ? Pourtant ds ton tableau il y en a 5... que tu as mis en commentaire ici mais pas ds le tableau..; IPV4 et geographi ???"} 

The first concern is the volume exchanged with IPv4 protocol. % \rouge{IPv4}\\
%metric is the \textit{volume of IPv4 and IPv6} traffic exchange. We know that IPv6 is the future of the Internet. If the traffic have none IPv6 makr, that means Reunion Island traffic have missed the train for the deployment for the version 6 of IP in the region. 
In this volume, we would make a distinction in terms of destination. We divide the collected traffic in 3 categories:
\begin {enumerate}
\item The first one is traffic in \emph{Local Area Network (LAN)} and concerned only traffic inside the intranet of collected points.
\item The second one can be associated to traffic in \emph{Metropolitan Area Network (MAN)}. With the presence of an \emph{Internet eXchange Point (IXP)}, we will, in this case, consider traffic intra-Reunion Island.
\item The last case is \emph{Wide Area Network (WAN)} traffic and identified traffic going outside of the Island. 
\end{enumerate}
The analyze of destinations included in the \emph{WAN} part would give us some indications about the location of the information needed by the user.
% We would also look the volume of traffic passing through the submarine cable to \textit{worldwide destination}, in comparison with the \textit{local traffic}. An analysis of the \textit{Internet access technologies} would give us an indication about the mobility of habits of the user. This metric would be compare to~\url{http://gs.statcounter.com/}. 
Our second metric we have focused on is the traffic's volume of different \textit{protocols of transport}, like TCP and UDP. This would inform us of the most used protocol and if the services use the right protocol.

Lastly, the number of the \textit{destination ports} will give us an overview of services that have been used. 
%We also look on the percentage on the \textit{geographical position of the destination}. We would see if the traffic have a preference for a country or a continent. The spread of the type of traffic, if its \textit{interactive or elastic} show us the kind of traffic consumed by the users.\\
\subsection{Protocol performance}
\label{subsec:perf}
\begin{figure}[ht!]
\centering
\includegraphics[width=0.48\textwidth]{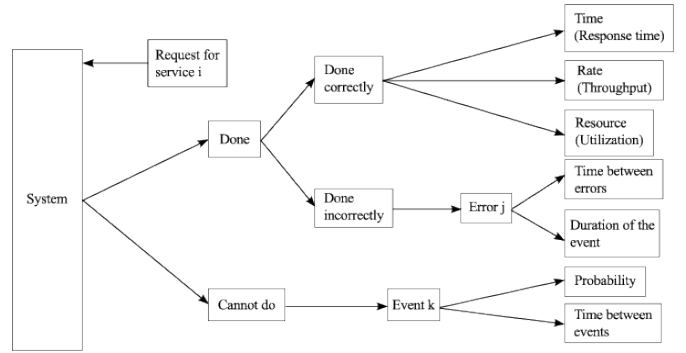}
\caption{Functioning of a request system}
\label{fig:jain}
\end{figure}

%% Pourquoi TCP ? Est-ce que TCP est le protocole le plus représentatif ?
%% Pourquoi ces métriques. En quoi sont-elles spéciales à étudier ?
The TCP is the principal protocol of transport used around the world with heavy-dependence on delay. Studying its performance is equivalent to study Internet performance. 
%Despite the results of the protocol transport spread's\yassine{CP}, we will analyze the performance of TCP. The \emph{Transmission Control Protocol (TCP)} is a transport protocol with heavy-dependence on delay. 
% In~\cite{Noordally2016}, the authors already have stated that Reunion Island has a specific%\yassine{idem}
In~\cite{Jain1991}, the author has 
%\yassine{Plusieurs auteurs ou un seul ? Rajouter un s ou bien mettre has} 
explained the functioning of a system based on requests. The figure~\ref{fig:jain} has been taken from his book. We used it for choosing which TCP metrics should be used.
The first and the second metrics are the RTT and the throughput. 
Some event could be easily identified, like packets losses and retransmits.\\
%relationship with delay and path. Studying some metrics of TCP, like losses, retransmits or congestion events\yassine{"il manque un bout de phrase... is important to understand the situation}. Different delay, like \emph{Round-Trip-Time (RTT)} or end-to-end delay will be measured.
% As presented in~\cite{Lan2006}, a new characterization of the flow has been proposed. We will use it to see the traffic in terms of flows. 
We define a flow as a bi-directional series of IP packet with the same source and destination address, port number and protocol number.

\subsection{Summary}
The table~\ref{table:metrics} resume all the metrics choose for the experiment.
\begin{table}[ht!]
	\centering
	\caption{Metric summary}
	\begin{tabular}{|p{0.1\textwidth}|p{0.1\textwidth}|p{0.1\textwidth}|}
\hline
\multirow{7}{0.1\textwidth}{\textbf{Traffic Supervision}}  & \multirow{3}{0.1\textwidth}{Traffic characterization}        & LAN              \\ \cline{3-3} 
                        &                     & \textit{MAN}          \\ \cline{3-3} 
                        &                     & WAN              \\ \cline{2-3} 
                        & \multicolumn{2}{c|}{Protocol Transport}                 \\ \cline{2-3} 
                        & \multicolumn{2}{c|}{Geographic Destination}               \\ \cline{2-3} 
                        & \multicolumn{2}{c|}{Services Used}                    \\ \cline{2-3} 
                        & \multicolumn{2}{c|}{IPv4 volume exchanged}                \\ \hline
\multirow{11}{0.1\textwidth}{\textbf{Protocol Performance}} & \multirow{8}{0.1\textwidth}{TCP performance}    & Congestion Event        \\ \cline{3-3} 
                        &                     & Losses             \\ \cline{3-3} 
                        &                     & Retransmissions        \\ \cline{3-3} 
                        &                     & Out of Order        \\ \cline{3-3} 
                        &                     & Windows            \\ \cline{3-3} 
                        &                     & Flags             \\ \cline{3-3} 
                        &                     & Connection Established Problem \\ \cline{3-3} 
                        &                     & Delay             \\ \cline{3-3} 
                        &                     & Throughput           \\ \cline{2-3} 
                        & \multirow{3}{0.1\textwidth}{Flows characterization} & Size              \\ \cline{3-3} 
                        &                     & Duration            \\ \cline{3-3} 
                        &                     & Rate              \\ \hline
\end{tabular}
   \label{table:metrics}
\end{table}

%-------------------------------------------------------
\section{Tools involved}
\label{sec:tools}
%% Outils utilisés pour les differentes phases du protocole de capture et d'analyze.

\subsection{TCPDump}
\label{subsec:tcpdump}
TCPDump~\cite{TCPDump} is a tool developed in 1989 by Jacobson, using the PCAP library~\cite{PCAP2001} to capture and analyze the Internet flows. This tool is the first tool created for this kind of metrology and allows the writing and reading of files in pcap format.

\subsection{Anonymization}
In agreement with the French law~\cite{LIL1978}, we need to anonymize the data obtained before the analysis.
The article~\cite{Minshall2005tcpdpriv} explain that an Internet packet contains private information and need to be removed before analysis. The author describes a tool to anonymize confidential information.
%In~\cite{Minshall2005tcpdpriv}, the author\yassine{"Tu peux dire directement "L'article cite.... explique que" "} presents \emph{TCPdpriv}, a tool for anonymize \yassine{"à revoir pluriel ou pas sans the mais avec s"} the confidential information. 
But in~\cite{Xu2002}, Xu explains %\yassine{"trouv eun synonyme... explain show put forward etc... y a trop de present dans ton texte"} 
some lack %\yassine{"pas le bon mot"} 
in \emph{TCPdriv} and proposes a new solution called~\emph{Crypto-pan}. \emph{Crypto-pan} is the most anonymized algorithm used actually. This program used the prefix-preserving system, which means that if two IP address have k-bits in common, their anonymized address keep k-bits in common.

\subsection{Analysis}
As said in~\ref{subsec:tcpdump}, PCAP is a library with several tools for traces analysis. One of the most well-know tool is Wireshark~\cite{chappell2010wireshark}. Wireshark can be used with a graphic-interface. But this tool is not adapted to our need, i.e. it doesn't take into account the geolocation of an IP address or it can't anonymize the raw data. %information about the weight of the collected data.
%presents some missing, like geolocation or information about volume\yassine{"reformuler... is not perfect, ie it doesn't take into account...}. 
To fill this gap, %avoid this missing\yassine{utilise ce site (combler une lacune) : http://www.wordreference.com/fren/combler}
 we have developed our own tool to obtain information about all metrics stated in section~\ref{sec:metrics}. This tool, developed in C, includes \emph{Crypto-pan} concept and focuses only on the metrics we have selected.
% This tool is available on the website of the author.\footnote{\url{http://lim.univ-reunion.fr/staff/rnoordal/media/tool_analysis.tar.gz}} \yassine{Ce lien n'est pas encore valide. Il le sera une fois que l'outil sera bien developpe, mais accessible lors du rendu.}

%-----------------------------------------------------------------
\section{Methodology}
\label{sec:methodo}
%% Protocole de mesure simplifiée et épurée de données non importantes.

% We chose to study the Internet performance and characteristics of Reunion Island. To do so we collected \emph{tcpdump} traces in the near point of the connectivity before the submarine cable (section~\ref{subsec:run}). To compare the Reunion Island connectivity, we also analyze traces from \emph{Center for Applied Internet Data Analysis (CAIDA)}.

% \subsection{Reunion Island}
% \label{subsec:run}
%In~\cite{Noordally2016}, the authors have illustrated a particular routing behavior with 97\% of the data goes throw France before/after joining/left Reunion Island.
The first study on Reunion Island connectivity was realized with the help of the \emph{Internet Service Provider} (ISP) was made in 2013 by~\cite{Vergoz2013}. In this paper, the authors have used active metrology to analyze the characteristics of the Reunion Island's Internet. To identify the performance of TCP, a better way is the listening of traffic and to analyze what happens in real-time or later. ISP information about their performances are very sensitive data, and this is the reason why they don't want to share it. 
To respect their publishing data politics, we have made our own measurement infrastructure. Deployed in the \emph{Computer Sciences and Mathematics Laboratory (LIM)} of Reunion Island University, we have collected data traffic from students and researchers. 
% LIM have an fiber-access to the Internet with a capacity of 500 Mbits/s and used only IPv4 address.
The collected data have been realized between one and 23 hours per day with port-mirroring protocol~\cite{Foschiano2015}.
The start and the duration of each measurement have been randomized. Two measurements have a maximal distance of 24 hours. We are fully aware that our collection is not representative of Reunion Island's traffic, but still a good set of tests for our analyze tools.%good sample of kind traffic could be generated on the Island.

% \subsection{CAIDA}
% \label{subsec:usa}
% Our second dataset was obtained by \emph{Center for Applied Internet Data Analysis (CAIDA)}\cite{Caida}. Since 2008, CAIDA have been created a public dataset of passive measurement. These traces have been collected from the data-center of Equinix located in Chicago. This dataset have only be available on demand\footnote{\url{http://www.caida.org/data/passive/passive_dataset_request.xml}}. In this dataset, we have the choice in the A-way or the B-way. As for our measurement, each traces was be collected during one our. 
% We have selected the most recently published data to compare the data. We will used CAIDA traces as the "standard".

%-----------------------------------------------------------------
\section{Results}
\label{sec:results}
%% Les differents resultats obtenus sur les metriques étudiées. Il va falloir être succint mais percutant car pas beaucoup d'espace. Necessite de justifier l'ensemble des valeurs obtenues. Pensez également a essayer de comparer avec d'autres resultats obtenues ailleurs si possible.
The results illustrated the output obtained by our analysis tool. It can be divided in two main parts. In subsection~\ref{subsec:carac}, we have the results of the supervision of the data collected. The subsection~\ref{subsec:tcp} concerns the analysis of the characteristics of the flows and the performance of TCP.%\yassine{Pas logique... ca devrait être ca + flows non ???}.

\subsection{Traffic characteristics}
\label{subsec:carac}
The LIM laboratory and in fact all laboratories and faculties of our University are connected to the Internet through a fiber access with a bandwidth of 500 Mbits/s with RENATER as ISP. RENATER is the French network reserved for Universities and research centers. During the month, we have collected 72 Gigabytes of data. This data concerned only IPv4 because IPv6 is not available in Reunion Island University.\\
% At Reunion Island University, the users are only allowed to use IPv4 protocol. This restriction is due to the absence of the protocol of IPv6 in the Island. 
% \rouge{Manque valeur size pcap file}
%% Spread of the destination traffic in term of Internal, Local (MAN) or 
As we stated before, we have three kinds of networks that have been analyzed: LAN (local network), MAN (network flow in Reunion Island) and WAN (network flows outside Reunion Island). With the analyze of the IP address, we have spread the traffic into three categories. The table~\ref{table:lwman} shown the results obtained.
%plot the figure~\ref{table:lwman} which shown the distribution of this traffic. 
\begin{table}[ht!]
\centering
\caption{Distribution of the LIM traffic}
\label{table:lwman}
\begin{tabular}{|c|c|c|c|}
\hline
Traffic & LAN & MAN & WAN \\ \hline
Percent & 21.65 & 0 & 78.35 \\ \hline
\end{tabular}
\end{table}
The WAN part is the main traffic of the laboratory. That's mean the information needed by the user is mainly located outside of the University or Reunion Island. \\

The table~\ref{table:destination} represent the geographical distribution of the IP addresses only present in the WAN traffic and has been regrouped by continents. % We can notice that Asia, Europe and North America take near of 95\% of the LIM traffic.
\begin{table}[ht!]
\centering
\caption{Geographic distribution of WAN destination}
\label{table:destination}
\begin{tabular}{|c|c|c|c|c|}
\hline
Continent & Africa & Asia & Europe & North America \\ \hline
Percent & 1.31 & 5.26 & 28.95 & 64.47 \\ \hline
\end{tabular}
\end{table}
We can notice that two continents, Oceania and South America, are not present in our data. It is surprising because the Laboratory has some partnerships present in Australia, for example.

% \begin{figure*}[ht!]
% % 	\begin{subfigure}[t]{\columnwidth}
% % 		\includegraphics[width=0.4\textwidth]{traffic.eps}
% % 		\caption{Distribution of the LIM traffic}
% % 		\label{curve:lwman}
% %   \end{subfigure}
% %   ~
% 	\begin{subfigure}[t]{\columnwidth}
% 		\includegraphics[width=0.98\columnwidth]{continent.eps}
% 		\caption{Geographic distribution of WAN destination}
% 		\label{curve:destination}
%   \end{subfigure}
%   ~
%   \begin{subfigure}[t]{\columnwidth}
%   	\includegraphics[width=0.95\columnwidth]{protocoles.eps}
% 		\caption{Spread of the transport protocol}
% 		\label{curve:protocol}
%   \end{subfigure}
% \end{figure*}
%% Est-ce que l'on parle des erreurs de géolocalisation ?

%% Description of the figure of the distribution LIM traffic.
The table~\ref{table:proto} represent the distribution of the protocol of transport from the Computer Science Laboratory by analyzing each packet captured. The first remarkable result is the presence of \emph{Internet Control Message Protocol (ICMP)} packet. %The presence of $5.72\%$ of ICMP packets can easily be explained by the experimentation of~\cite{Noordally2016}. \yassine{Tu dverais mettre plutot: 
The presence of $5.72\%$ of ICMP packets is due to a measurement delay campaign as it has been explained in~\cite{Noordally2016}.
%}
\begin{table}[ht!]
\centering
\caption{Spread of the transport protocol}
\label{table:proto}
\begin{tabular}{|c|c|c|c|l|}
\hline
Protocol & ICMP & IGMP & TCP & UDP \\ \hline
Percent & 5.72 & 0.01 & 92.49 & 1.78 \\ \hline
\end{tabular}
\end{table}
%% UDP low. We can expect that the % is higher due to the strong presence of Chrome as Web Browser in Reunion Island and the utilization of the QUIC flag in Chrome.
The percentage of UDP is low if we are taking into consideration the fact that Google Chrome is the most used browser in Reunion Island as presented on a website\footnote{\url{http://gs.statcounter.com/#browser-RE-monthly-201511-201612}}. This web browser has an option call \emph{QUIC} which increase the performance of the browser by using \emph{HyperText Transport Protocol (HTTP)} over \emph{User Datagram Protocol (UDP)}~\cite{Carlucci2015}.

% \yassine{Tu devrais pitet dire que bcp de ports sont bloqué par le parefeu univ, d'où cette présence énorme de TCP... dans la vie réelle c'est pas comme ça (VOD, Jeux vidéo, Skype, etc.). D'ailleurs je trouve bizarre ce UDP. A mon avis, c'est pas chrome mais plutôt la visio ou autre non ?}

Needless to say, TCP have the higher percentage between the different protocol. One reason can be that Reunion Island's University have set up some rules to limit the traffic generated by the users and closed some port. For example, Skype is prohibited as call-conference and the majority of TCP ports except the common like 22 for \emph{Secure-Shell (SSH)} or 25 for \emph{Simple Mail Transfer Protocol (SMTP)} are closed.% if you asked for specific reason are closed. 
The table~\ref{tab:service} illustrate the analyze of TCP ports. 
%The first analyze of TCP packets concern the service use\yassine{CP}. The figure~\ref{curve:service} represent the distribution of services used by identifying the port number. As explained before, the traffic are filtered by the University and some port are closed. 
% \begin{figure}[ht!]
% \includegraphics[width=0.4\textwidth]{services.eps}
% \caption{Distribution of the service used by TCP}
% \label{curve:service}
% \end{figure}
\begin{table}[ht!]
\centering
\caption{Distribution of the service used by TCP}
\label{tab:service}
\begin{tabular}{|p{0.1\columnwidth}|p{0.05\columnwidth}|p{0.05\columnwidth}|p{0.05\columnwidth}|p{0.14\columnwidth}|p{0.06\columnwidth}|p{0.08\columnwidth}|p{0.08\columnwidth}|}
\hline
Service & SSH & DNS & Mail & Non indentified & Other & HTTPS & HTTP \\ \hline
Percent & 0.02 & 1.63 & 4.91 & 7.21 & 9.45 & 32.97 & 43.79 \\ \hline
\end{tabular}
\end{table}
The item \emph{Non identified} regroup all port number without any matching in the file 'service' located on the analysis server. \emph{Other} identify protocols with a percentage lower than 1\%. The presence of SSH is explained by its frequently use to push data on the web-server of the laboratory or to connect to distant servers.
The \emph{Mail} item regroup five protocols consecrate to send or receive email, like \emph{Post Office Protocol (POP)}, SMTP or \emph{Internet Message Access Protocol (IMAP)} and their secured version, POP3s, IMAPs. But these protocols are only present if you use an email application like Thunderbird or Outlook. If you connected to Gmail and use the web interface, it will be considered as HTTP or HTTPS.

%More and more website used the protocol HTTPS to share their information, like Google. The huge presence of this service is not surprising. HTTP are also used but with lower percent. SSH and Mail are also present but in little part. The first is used to connected to University server to work on. The mail 
%This is the reason why we did a strong analyze of the performance of this protocol in order to explain the sensation of slowness described by users.
The huge presence of TCP is the reason why we will make an %did an % strong
analyze of the performance of this protocol in order to explain the sensation of slowness described by users.

\subsection{TCP Performance}
\label{subsec:tcp}
\begin{figure}[ht!]
\begin{subfigure}[h]{\columnwidth}
\centering
\includegraphics[width=0.98\columnwidth]{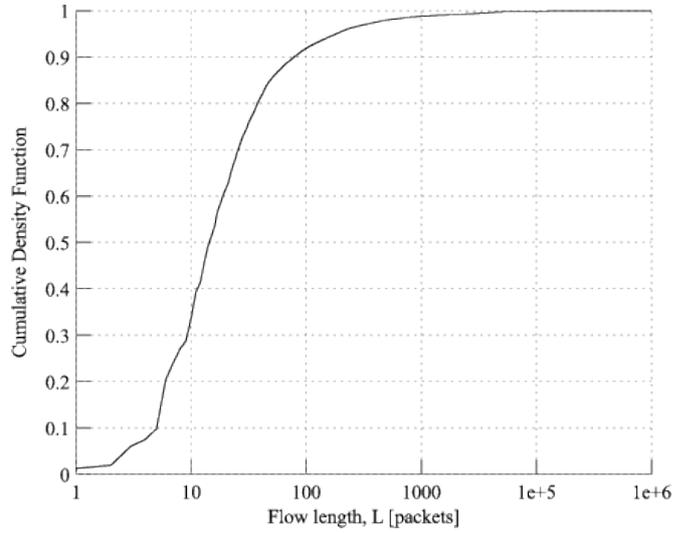}
\caption{Length distribution}
\label{curve:flow_length}
\end{subfigure}
\begin{subfigure}[h]{\columnwidth}
\centering
\includegraphics[width=0.98\columnwidth]{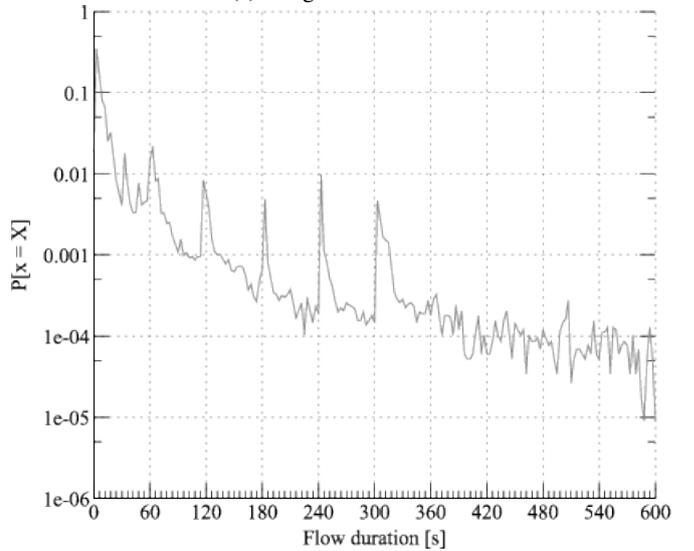}
\caption{Duration density}
\label{curve:flow_duration}
\end{subfigure}
\caption{TCP Flow}
\end{figure}

We have analyzed the different types of flows present in the data analysis.
The figure~\ref{curve:flow_length} represent the Cumulative Density Function (CDF) for the flow length L in packets. 
We can notice that more than 90\% of our flows are composed of fewer than 100 packets. These flows represent $94.45$\% of the volume exchange in bytes. That supposed that the packet exchanged are heavy in terms of data.
% With the wrong used of the TCP port by the users, and the rules made by the Reunion Island University, it is difficult to identify the link of the $5.55$\% remaining in term of services. We can only supposed that could be interactive traffic, like Skype, or some downloaded files with HTTP.

The figure~\ref{curve:flow_duration} represent the PDF % \yassine{ je crois que tu as utilisé qqpart avant sans le définir... à revoir juste avant le III}
for the flow duration in seconds. The bin used for the PDF is 3s. We can see that the first peak represents the Recovery Time Objective (RTO) of TCP. These flows represent only $3.5$\% of the volume exchange. Exceed the RTO, the probability of the duration decrease. We can notice that on the figure, every 60s, a peak are present. The flows exceeding 300s (5~min) represent 59\% of the volume exchange. That means more the flows carry on, more the volume exchange increase.

\begin{figure}[ht]
\centering
\includegraphics[width=0.98\columnwidth]{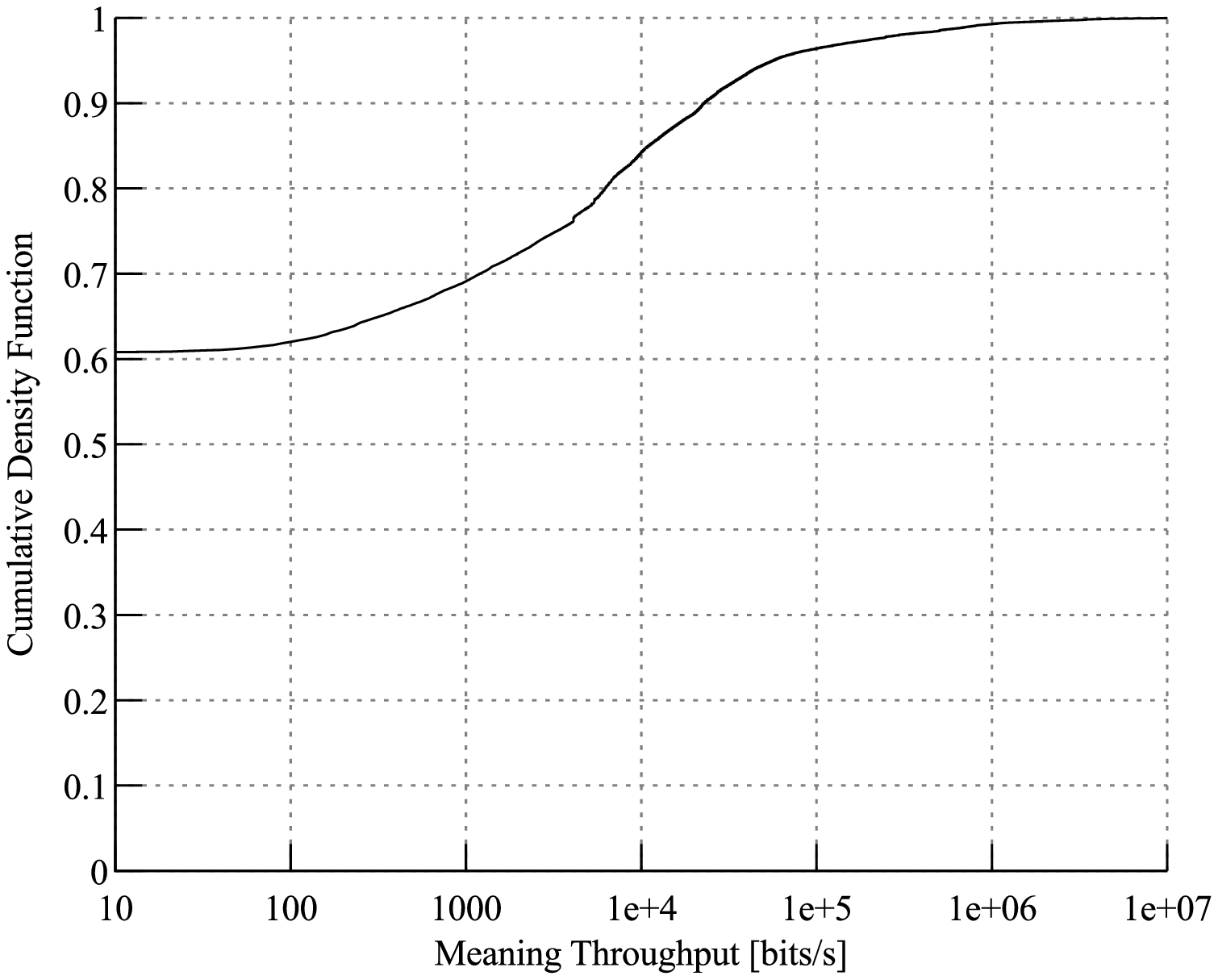}
\caption{}
\label{curve:flow_rate}
\end{figure}

With the figure~\ref{curve:flow_rate}, we can analyze the meaning throughput of each flow.
We can notice that 60\% of our traffic have a meaning throughput less or equal than $10$ bits per seconds, but this majority represent only less than $10^{-3}\%$ of the volume exchange. That's mean the flows are very light in terms of bytes exchange, and could be composed by \emph{Keep-Alive} packet. 

TCP is a protocol relying on delay but Reunion Island have a specific behavior on this parameter~\cite{Noordally2016}. A supposition we can do is that Reunion Island's traffic is very sensitive to congestion events. We have two ways to confirm it. %identify congestion events:
\begin{enumerate}
\item The first one is to identify the different flag of the flow and see if the \emph{Explicit Congestion Notification(ECN)}~\cite{rfc3168} and the \emph{Congestion Window Reduce (CWR)} flags are put to one.
\item The second way is to identify the reduction of TCP's windows and see what happens after. 
\end{enumerate}
With the first way of Congestion Event identification, none event was found. That means the different flag for the management of congestion event didn't rise up. The authors of~\cite{trammell2015enabling} shown that the number of servers which answered to ENC negotiation has increased. The missing of mark on the client side could be explained by two hypothesis.
%\yassine{Il manque un bout de phrase... cest la suite ?}. Two hypothesis could be claimed. 
The first one is the obsolescence of the computer used in the Laboratory. But the machine used by our colleagues have been bought recently (current 2015 for someone). The second hypothesis is the remove of the flag by the equipment. 
In both hypotheses, the absence of the ECN flags are harmful for the TCP performance. In~\cite{mellia03}, the authors have made some traffic simulation about TCP performance and packet marked. The results shown that the performance of TCP will increase when the packets have been marked.
Without these flags, the Congestion event is more difficult to identify. 
The second way is more difficult to set up. We need to identify the windows reduced and see if we could identify some Lost Packet around this event. We have identified $152,230$ packets which indicate a window reduction. It's represented only $3\times10^{-3}$ percent of the packets exchange. But with only $232,789$ flows, a flow has $65$\% of chance to include a packet marked \emph{Windows reduce}.
The tabular~\ref{table:pkts} represent the percent of packets retransmits or losses.
\begin{table}[ht!]
\centering
\caption{Packets spreading}
\label{table:pkts}
\begin{tabular}{|p{0.1\columnwidth}|p{0.2\columnwidth}|p{0.2\columnwidth}|p{0.1\columnwidth}|p{0.15\columnwidth}|}
\hline
Packets & Fast Retransmit & Spurious Retransmit & Lost Packet & None retransmit or lost \\ 
\hline
Percent & $8.3\times10^{-3}$ & $9.7\times10^{-3}$ & $18\times10^{-3}$ & $99.96$ \\ 
\hline
\end{tabular}
\end{table}
We can see that the percent of packets retransmits or losses are like crumbs, with less than 1\%. The percent of these errors are less than TCP could produce in normal condition.

%% Petit paragraphe explicatif sur la dégradation des performances de TCP (si c'est le cas) à cause de la connectivité particulière de la Réunion à Internet.
%% submarine cable map and one reason of the degraded TCP performance

\section{Related works}
\label{sec:related}
% Related work : Essayer de trouver plus d'article sur les performances dans la zone africaine et utiliser Bischof2015 comme point de repère si possible. Sinon chercher d'autres articles sur des études de performances de l'Internet dans le monde. 
A description of Cuba Internet's connectivity has been done in~\cite{Bischof2015}. The authors have shown that, despite a similar Internet access than Reunion Island, Cuba have an Internet latency near to the general case. This general case was presented in~\cite{Krajsa2011}, where a relation between delay and geographical distance have been made. The authors of~\cite{Noordally2016} have proved that Reunion Island is a specific case. In analyzing the Internet performance of the African continent, several studies have been made.
%The performance of Africa Internet is poor.\yassine{CP le role de cette phrase}
In 2001,~\cite{Johnson2011} have highlighted that the delay is very high and the HTTP response very low in rural area of Africa. In comparison with us, their flows are less longer, with 60\% of their traffic have a duration lower than 2 seconds.
% \rouge{comparison of flow duration between us and them}
The presence of flow durations higher than 10 minutes is a little fraction (0.47\%) but are still present in the analyze. 
Another study has been done in a rural area of the African continent, more specifically in Zambia. The study has been done in 2005 by the author of~\cite{zheleva2015internet} after an upgrade of the bandwidth. However, the meaning delay to reach services stayed higher than elsewhere.
%A study of the evolution of the traffic in rural Zambia after an upgrade of the bandwidth has been made in~\cite{zheleva2015Internet}. The authors have found better performance for the users, but in general\yassine{j'aime pas ca... faut trouver autre chose mais je trouve pas ;)} for the African continent, the meaning delay to reach services is higher than elsewhere. 

A recent research about the presence of services in Africa has been explained in~\cite{fanou2016pushing}. The author has explained that despite the presence of servers in the continent, most of the traffic continues to go to America. It is a very similar situation as in Reunion Island with France servers.

% There are several ways to measure the RTT with passive measurement. In~\cite{Jiang2002}, the authors proposes two different estimation of RTT, using SYN-ACK and Slow-Start. Contrary to~\cite{jaiswal2004}, which used the \yassine{CP}

% Contrary to other country like Japan, \yassine{CP}

\section{Conclusion}
\label{sec:conclusion}
Making a campaign of passive metrology is the second step for a better knowledge of the reasons why the Internet is degraded in some regions. Reunion Island has a specific connectivity to the Internet, with two submarine cables and a routing rules which made that the majority of the traffic going through France. 
Our study shows that up to 78\% of the LIM traffic is to WAN and TCP is the most important transport protocol used in Reunion Island. Studying its performance is important to identify the default of the Reunion Island connectivity but the filter set up by the University impact our results. 
The analysis shows that the feeling of slowness didn't come %\yassine{Je comprends pas le sens de la phrase} 
from the performance of TCP but only due to the delay and the routing rules. %\yassine{manque la fin de phrase}
%the and its performance are not so degraded in term of losses. The filter set up by the network department of the University impact the services used in the University and our results. 
%% Future work : Travailler sur les délais dans la ZOI et après sur les performances de TCP dans la ZOI. Trouver d'autres zones dans le monde qui ont des performances Internet similaires à La réunion.
%In the future, we would make a comparison between Reunion Island's Internet characteristics and the other Islands in the Indian Ocean, like Mauritius, Madagascar and Seychelles for example. We would also try to identify other regions where the Internet access has some similarities with Reunion Island.
% In addition to the performance, we will supervise Reunion Island's Internet.
In the future, we will contact the ISP working on Reunion Island and try to make an analysis of TCP performance of the Isl. In addition to the performance, we will supervise Reunion Island's Internet. We would also try to exchange information with the ISP present in the Indian Ocean countries, like Madagascar, Mauritius or Seychelles to make a comparison in the region.
% \section*{Acknowledgments}
% The authors would like to thanks all people would have help for the development of the tool analysis.
%% \bibliography{metrology}
%% \bibliographystyle{IEEEtran}

\end{document}